\documentclass[12pt,preprint]{aastex}

\usepackage[percent]{overpic}
\usepackage{graphicx}
\usepackage{natbib}

\newcommand{\kms}{\hbox{km~s$^{-1}$}}
\newcommand{\kmsmpc}{\hbox{km~s$^{-1}$~Mpc$^{-1}$}}

\begin{document}

\title{A Potential Recoiling Supermassive Black Hole CXO J101527.2+625911}

\author{
D.-C. Kim\altaffilmark{1}, 
Ilsang Yoon\altaffilmark{1},
G. C. Privon\altaffilmark{2},
A. S. Evans\altaffilmark{1, 3},
D. Harvey\altaffilmark{4},
S. Stierwalt\altaffilmark{1}, \&
Ji Hoon Kim\altaffilmark{5}
}
\altaffiltext{1}{National Radio Astronomy Observatory, 520 Edgemont Road,
Charlottesville, VA 22903: dkim@nrao.edu, aevens@nrao.edu, sstierwa@nrao.edu}

\altaffiltext{2}{Instituto de Astrof{\' i}sica, Facultad de F{\' i}sica, Pontificia Universidad Cat{\' o}lica de Chile,
Avda. Vicuna Mackenna 4860, Santiago, Chile, Codigo Postal: 8970117}
\altaffiltext{3}{Department of Astronomy, 530 McCormick Rd., University of Virginia, Charlottesville, VA 22904}
\altaffiltext{4}{Laboratoire d’Astrophysique, EPFL, Observatoire de Sauverny, Chemin des Maillettes, 51, Versoix CH-1290, Suisse}
\altaffiltext{5}{Subaru Telescope, National Astronomical Observatory of Japan, 650 North A'ohoku Place, Hilo, HI 96720}

\begin{abstract}
We have carried out a systematic search for recoiling supermassive black holes (rSMBH) using the Chandra Source and SDSS Cross Matched Catalog.
From the survey, we have detected a potential rSMBH, 'CXO J101527.2+625911' at z=0.3504.
The CXO J101527.2+625911 has a spatially offset (1.26$\pm$0.05 kpc) active SMBH and 
kinematically offset broad emission lines (175$\pm$25 km s$^{\rm -1}$ relative to systemic velocity). 
The observed spatial and velocity offsets suggest this galaxy could be a rSMBH, but
we also have considered a possibility of dual SMBH scenario.
The column density towards the galaxy center was found to be Compton thin, but no X-ray source was detected.
The non-detection of the X-ray source in the nucleus
suggests either there is no obscured actively accreting SMBH, or there exists an SMBH but has a low accretion rate (i.e. low-luminosity AGN (LLAGN)).
The possibility of the LLAGN was investigated and found to be unlikely based on the H$\alpha$ luminosity, radio power, and kinematic arguments. 
This, along with the null detection of X-ray source in the nucleus supports our hypothesis that 
the CXO J101527.2+625911 is a rSMBH.
Our GALFIT analysis shows the host galaxy to be a bulge-dominated elliptical.
The weak morphological disturbance and small spatial and velocity offsets suggest that 
CXO J101527.2+625911 could be in the final stage of merging process and about to turn into a normal elliptical galaxy.

\end{abstract}


\section{Introduction}

Galaxy interactions play an important role in their evolution;
they enhance starburst activity (Larson \& Tinsley 1978; Joseph et al. 1984; Sanders et al. 1988), 
induce starburst and active galactic nuclei (AGN) feedback (Lehnert \& Heckman 1996; Heckman et al. 2000; Rupke et al. 2002), 
enrich the intergalactic medium with outflows (Nath \& Trentham 1997; Scannapieco et al. 2002),
and aid the formation and growth of stellar bulges and supermassive black holes (SMBHs) (Kormendy \& Richstone 1995; Magorrian et al. 1998; Gebhardt et al. 2000; Ferrarese \& Merritt 2000; Tremaine et al. 2002; Marconi \& Hunt 2003; Hopkins et al. 2005; Kormendy \& Ho 2013).
When two SMBHs coalesce at the final stage of a galaxy interaction,
{\it a merged SMBH} can recoil from the host galaxy due to anisotropic emission of gravitational waves (Peres 1962).
Recent simulations of merging black holes predict that the merged SMBH can attain a kick velocity of
a few hundred to a few thousands km s$^{-1}$ depending on mass ratios, spin magnitudes,
and spin orientations of the merging SMBHs (Campanelli et al. 2007; Schnittman 2007; Baker et al. 2008; Lousto \& Zlochower 2011; Blecha et al. 2016).
If the merging SMBHs are of equal mass and highly-spinning, and their spins are aligned along the orbital plane (superkick configuration), 
recoil velocities as large as 4000 km s$^{-1}$ (Campanelli et al. 2007) to 5000 km s$^{-1}$ (Lousto \& Zlochower 2011) can be reached
and the recoiling SMBH (hereafter rSMBH) will eventually escape from the host galaxy (i.e., Merritt et al. 2004).

The recoiling supermassive black hole (rSMBH) carries along with it the broad-line region (BLR) and leaves the stellar nucleus behind;
it can be observable for tens of Myr as an offset AGN (Madau \& Quataert 2004; Loeb 2007; Blecha et al. 2011, 2016).
Thus, we expect two observational characteristics:
i) the rSMBH could be observed spatially offset with respect to the stellar center of the host galaxy, and
ii) the broad emission lines could have a measurable velocity offset with respect to the systemic velocity.
So far, there have been more than a dozen reports on the rSMBH candidates.
Some examples of the rSMBH candidates are 
SDSS J092712.65+294344.0 (Komossa, Zhou \& Lu 2008),
SDSS J105041.35+345631.3 (Shields et al. 2009),
M87 (Batcheldor et al. 2010),
QSO E1821+643 (Robinson et al. 2010),
CXOC J100043.1+020637 (Civano et al. 2010),
CXO J122518.6+144545 (Jonker et al. 2010),
a half-dozen SDSS QSOs (Eracleous et al. 2012),
10 nearby core elliptical galaxies (Lena et al. 2014),
NGC 3115 (Menezes, Steiner, \& Ricci 2014),
five SDSS AGNs (Comerford et al. 2015), and 
twenty six SDSS QSOs (Kim et al. 2016).
Except for CXOC J100043.1+020637, most of the rSMBH candidates to date either have a spatial offset or a velocity offset, but not both. 
The CXOC J100043.1+020637 has two compact sources separated by $\sim$2.5 kpc and has a velocity offset of $\sim$1200 $\kms$ (Civano et al. 2010).
Chandra observations find that the south-eastern source that has a point-like morphology typical of a bright
AGN is responsible for the whole X-ray emission in this system (Civano et al. 2012).
The north-western source has a more extended profile in the optical band with a scale length of $\sim$0.5 kpc.
Recent 3 GHz Karl G. Jansky Very Large Array (VLA) observations find that the entire observed 3 GHz radio emission can be
associated with the south-eastern nucleus (Novak et al. 2015).
Their finding preferred rSMBH picture but cannot rule out the presence of an obscured and radio-quiet SMBH in the north-western source (i.e. Blecha et al. 2013; Wrobel, Comerford, \& Middleberg 2014).

Here, we report an analysis of optical imaging and spectroscopy, as well as X-ray data, of 'CXO J191527.2+625911' and show that it is one of the best rSMBH candidates to date.
The paper is divided into four sections. 
In Section 2, detection of the CXO J101527.2+625911 from the systematic search for the rSMBH is described. 
The discussion is presented in Sections 3 and a summary of the paper is presented in Sections 4. 
Throughout this paper, the cosmology H$_0$ = 70 \kmsmpc, $\Omega_M$ = 0.3, and $\Omega_\Lambda$ = 0.7 is adopted.
 
\section{Detection of a Potential rSMBH CXO J101527.2+625911}

We have conducted a systematic search for rSMBHs from
Chandra Source Catalogs (CSC) - Sloan Digital Sky Survey (SDSS) Cross-Match Catalog (Evans et al. 2010; Rots \& Budav{\' a}ri 2011).
The CSC-SDSS Cross-Match Catalog (CSC-SDSS CC) contains a total of 19,275 sources 
of which SDSS images for all and SDSS spectra for a significant part of the sources exist,
enabling us to identify both spatial and velocity offsets of the rSMBH candidates.
Another advantage of using the CSC-SDSS CC is that all entries are X-ray sources of which many of them are AGNs. 
In addition to the SDSS images, we have searched the Hubble Space Telescope (HST) for images which contain sources in the CSC-SDSS CC. 
The initial selection of possible images containing the rSMBH candidate was made by searching the HST archive for WFPC1/2, NICMOS, ACS and WFC3 images 
with central coordinates within a 2\farcm5 radius of the CSC-SDSS CC position for each source. 
This selection criteria returned matches for 3,873 sources. 
Each HST image was then examined to see if it contained the optical/near-infrared counterpart of the CSC-SDSS CC sources. 
In the end, 2,542 of the sources have HST imaging data.  
The next step in the selection process was to examine the HST and SDSS images in order to determine the morphology of the host galaxy. 
Host galaxies with overlapping nuclei or double nuclei within a half radius of a single galaxy were selected. 
However, if the galaxies are detached doubles, apparently interacting, or strongly disturbed, we have excluded them since we cannot accurately determine the center of these galaxies. 
For the selected rSMBH candidates, we then inspected the SDSS spectra. 
In cases in which these sources have broad-lines in the spectra and have only a single set of narrow emission lines, 
we performed a spectral decomposition to see if any broad-line velocity offset relative to systemic velocity was observed. 
For the spectral decomposition, we used the IRAF/Specfit package with 3 component fits in the H${\alpha}$ region: 
i) power-law continuum, ii) broad emission line, and iii) narrow emission line. 
We applied a single Lorentzian or a single Gaussian profile for the H${\alpha}$ broad-line component and a single Gaussian profile for the H${\alpha}$ and [N II] narrow line components.
The same Gaussian line widths were used for the narrow emission lines of H${\alpha}$ and [N II], and
a fixed value of $1/3$ was used for the [NII]6548 to [NII]6583 line ratio.
In this process, we excluded sources with two sets of narrow emission lines since they are most likely interacting galaxies or dual SMBHs. 

A potential rSMBH candidate, 'CXO J101527.2+625911 (z=0.3504)' was discovered in this process.
In the HST/ACS I-band (F775W) image (Treu et al. 2007) in Fig. 1a, the CXO J101527.2+625911 looks more like an E/S0 galaxy 
with tidal features on the east side (thick tidal tail) and west side (thin spiral-like structure).
In the center of the galaxy, two nuclei are clearly visible, one in the north and the other in the south.
The two nuclei are more prominent in the zoom-in image with contour plot overlaid (Fig. 1b):
a southern nucleus and a much brighter ($\sim6 \times$) northern nucleus. 
The position of the southern nucleus is near the center of the galaxy and the northern nucleus is offset from the center. 
In order to determine the relative positions of each nucleus within the host galaxy, 
we have performed an ellipse fitting.
Before the ellipse fitting, we have subtracted the northern nucleus component
since it is apparently offset from the center of galaxy and will produce a wrong centroid.
The ellipse fitting was done within the circled region in Fig. 1a using IRAF/Ellipse package 
(we excluded disturbed and asymmetric part of the galaxy in the outside of the circled region since it will produce wrong centroid).
Ellipse fitting produces center positions of the fitted ellipses as a function of radius.
We calculated positional offsets of the center of each ellipse with respect to the center of southern nucleus.
If the southern nucleus is the true host galaxy center, we will see a small or no offset.
The result is plotted in Fig. 1c where the magenta circle and green triangle represent a start and end point of the fitting, respectively.
The positional offsets range from 0 to 0\farcs1 and is less than the FWHM of ACS point spread function (PSF, 0\farcs1 to 0\farcs14).
This suggests that the southern nucleus is a center of host galaxy within a positional uncertainly of the ACS.
Hereafter, we will call the southern nucleus as a nucleus of host galaxy.

The absolute astrometry of the HST images is $\sim$ 0\farcs3 thus, we can not measure the positions of the two nuclei better than this limit.
To find a more accurate position, 
we have searched the large quasar reference frame (LQRF) catalog (Andrei et al. 2009) and 
matched a quasar position for our source.
This position and its positional uncertainty (0\farcs135) is marked with white dot and circle, respectively in Fig. 1b.
The magenta plus symbol and circle in the plot represent the position of CXO J101527.2+625911 from Chandra Source Catalog Ver. 2 and its positional uncertainty (0\farcs32), respectively. 
As shown in the image, the Chandra position is closer to the northern nucleus and the positional uncertainty circles of the LQRF and Chandra partly overlap each other.
This suggests that the northern nucleus is the X-ray source (hereafter an offset AGN or offset SMBH).
The projected nuclear separation between the offset AGN and the nucleus of the host galaxy is 0\farcs26$\pm$0\farcs01 corresponding to physical scale of 1.26$\pm0.05$ kpc. 

To estimate the recoil velocity, a spectral decomposition was performed for H$\beta$ line in the high S/N Keck Low Resolution Imaging Spectrometer (LRIS) spectra (resolution $\simeq$ 55 \kms) observed by Woo et al. (2006).
The H$\alpha$ line in the SDSS spectra could be used to measure recoil velocity,
but we decided to use the Keck LRIS spectra since the redshifted H$\alpha$ line lies on the edge of SDSS spectral coverage and contains noisy signals (Fig. 2a).
For the spectral decomposition, we used the same method ass we did for the H$\alpha$ line.
However, the [Fe II] component which is often found in the QSOs was not added in the fit 
since we do not see this line near the H${\beta}$.
The result of spectral decomposition is presented in Fig. 2b, where the black, cyan, blue, green, and red lines represent
data, power-law continuum, broad emission line, narrow emission lines, and model (sum of all fitting components), respectively.
Vertical dotted and dashed lines represent the line center of the H$\beta$ systemic velocity and the H$\beta$ broad-line, respectively
and the systemic velocity (z=0.3504) was measured from the low-ionization forbidden line [S II]$\lambda 6716$.
The H${\beta}$ broad-line is redshifted by 175$\pm$25 \kms\ relative to systemic velocity and its FWHM is 4200 \kms.

The radio emission from this galaxy was detected (S$_{1.4 GHz}$=1.6 mJy) from Faint Images of the Radio Sky at
Twenty centimetres (FIRST: Becker et al. 1995) VLA Sky Survey.
The radio luminosity calculated from L$_{1.4 GHz}$=4$\pi$D$_l^2$(1+z)$^{\alpha -1}$S$_{1.4 GHz}$ with luminosity distance D$_l$=1859 Mpc and spectra index $\alpha$=0.7 is 6.1$\times 10^{23}$ W Hz$^{-1}$ places this galaxy into radio-loud category ($>10^{23}$ W Hz$^{-1}$).
Its q-value (FIR to radio luminosity ratio, Helou et al. 1985) is 2.08 and similar to the mean value found in quasars but lower than that found in starburst galaxies (Mori{\' c} et al. 2010).
The X-ray luminosity of the CXO J101527.2+625911 is L$_{0.5-7.0 keV}=2.39\times 10^{43}$ erg s$^{-1}$ cm$^{-2}$ and about an order of larger than X-ray-selected broad-line AGNs in the similar redshift range (Suh et al. 2015).
The basic properties of the CXO J101527.2+625911 are listed in Table 1.


\section{Discussion}
  
\subsection{A rSMBH or a dual SMBHs?}

Though we have detected both spatial and velocity offsets in the CXO J101527.2+625911,
this system could be a dual SMBHs.
If both AGNs in a dual SMBHs are actively accreting, we will observe two sets of shifted narrow and broad lines in the spectra
since each AGN has its own BLR and narrow line region (NLR) and both AGNs will follow gravitational potential of the host galaxy.
If their NLRs are mixed together, we will still observe two sets of shifted broad lines and one set of narrow lines.
Detection of only the redshifted broad lines relative to narrow lines in the CXO J101527.2+625911 excludes two actively accreting dual SMBHs scenario.  
It could be possible that the SMBH in the nucleus is actively accreting, but obscured behind gas and dust.
X-ray observations will be one of the best methods to detect the obscured SMBHs.
From Chandra ACIS-S point source sensitivity limit (4 $\times 10^{-15}$ ergs cm$^2$ s$^{-1}$ in 10$^4$ sec of exposure time),
Compton thickness (Juneau et al. 2011) in the nucleus can be estimated.
If there exists an obscured SMBH in the nucleus of host galaxy, 
the [O III] flux we have measured will be a combination of emission from the offset SMBH and obscured SMBH.
If we assume about a half of the [O III] flux comes from the obscured SMBH, 
the Compton thickness in the nucleus will be Compton thin (log (L$_{\rm{X-ray}}$/L$_{\rm{[O III]}}$) = 0.70).
Even if all of the [O III] flux comes from the obscured SMBH, it is still Compton thin (log (L$_{\rm{X-ray}}$/L$_{\rm{[O III]}}$) = 0.40).
So, if there exists an obscured SMBH in the nucleus of host galaxy and is actively accreting, 
we could have detected the X-ray emission.

The non-detection of the X-ray source in the nucleus suggests 
there is no actively accreting SMBH, or there exists an SMBH but with a low accretion rate (i.e., low-luminosity AGN (LLAGN)).
The possible existence of the LLAGN in the nucleus was investigated, but was found unlikely based on the followings:
i) The typical H$\alpha$ luminosity and 5 GHz radio power of the LLAGNs are L$_{H\alpha}$=1.7$\times 10^{39}$ erg s$^{-1}$ and P$_{radio}=8.5\times 10^{19}$ W Hz$^{-1}$, respectively (Ho 2008),
whereas these quantities are more than 2 to 3 orders of magnitude larger in CXO J101527.2+625911 
(L$_{H\alpha}$=3.8$\times 10^{41}$ erg s$^{-1}$ and P$_{radio}=2.5\times 10^{23}$ W Hz$^{-1}$), and
ii) If there exists a LLAGN in the nucleus, the pair of SMBHs will move in a circular orbit with a rotation velocity of $\sim$500 \kms (assuming both black hole masses are comparable).
However, the measured velocity is only 175 km/s (it will be $\sim$260 \kms if we consider the inclination angle and the longitude of ascending node).
The above X-ray and LLAGN arguments favor a rSMBH scenario rather than a dual SMBHs one.

It is possible that the source identified as an offset AGN could instead be a bright background source.
If a background source is located beween z=0.3504 and z=1,
then the comoving volume extended by a solid angle of 0\farcs26 diameter is 1.3$\times 10^{-3}$ Mpc$^3$.
The space density of X-ray sources with L$_{0.5-7.0 keV}=2.39\times 10^{43}$ erg s$^{-1}$ cm$^{-2}$ at z=0.3504 is $5 \times 10^{-5}$ Mpc$^{-3}$ from the X-ray luminosity function of Silverman et al. (2008).
If the same source is placed at z=1.0 then the required X-ray luminosity is L$_{0.5-7.0 keV}=3.05\times 10^{44}$ erg s$^{-1}$ cm$^{-2}$ and its space density becomes $8 \times 10^{-6}$ Mpc$^{-3}$.
So, the overlap probability ranges from 6.5$\times 10^{-8}$ (for a background source at z=0.3504) to 1.1$\times 10^{-8}$ (z=1.0) and can be ignored.

We find that the narrow line ratios of [O III]/H$\beta$ and [N II]/H$\alpha$ are consistent with excitation by an AGN (log([O III]/H$\beta$)=1.24, log([N II]/H$\alpha$)=-0.10).
The average size (diameter) of the NLR in Seyfert 1 galaxies is about D=4.6 kpc (Bennert et al. 2006).
If we assume that the NLR in our source is simiilar to this, then the offset AGN is still within the NLR, but not near the center.
So, the offset AGN ionizes NLR gas in density-stratified NLR environment (ionizes more NLR gas in the center direction and less in the opposite direction).
We do not think an obscured (if exists) nuclear AGN ionized the NLR gas since
it is not detected both from the X-ray and the HST near-infrared J-band image even if the nucleus of galaxy is optically thin ($\tau_J$=0.25).



\subsection{Host Galaxy Type}

Modeling of the galaxy's morphology can provide information about the dynamical history of the system. 
We expect rSMBHs to be in systems with galaxy mergers in their past, 
and these events will likely be evident in the morphology and light profile of the remnant galaxy.
The fitting of the host galaxy requires a careful subtraction of the offset AGN.
To generate a realistic point spread function at the position of the AGN,
we create a TinyTim model of the ACS PSF for the F775W filter for each exposure.
Given that Hubble has a time dependent variability to the PSF due to thermal breathing,
we estimate the focus position of space telescope by measuring the shapes of the stars and 
compare them to 16 different focus positions and find the best fitting PSF model (Harvey et al. 2015).
With the wavelength and focus position for each exposure, a PSF is generated in pixel space 
and then combined at the same pixel scale as the data using the publicly available package AstroDrizzle.
Two dimensional galaxy fitting was performed with GALFIT 3.0 (Peng et al. 2010) using a composite model as shown in Fig. 1d: 
the PSF model (upper-left inset), bulge component (S{\' e}rsic index n=4, upper-right inset), and disk component (S{\' e}rsic index n=1, lower-left inset).
The fitted host galaxy turned out to be a bulge-dominated (log (Bulge/Disk) = 1.1) elliptical galaxy. 
In the residual image (lower-right inset), we see an artifact of imperfect PSF subtraction, 
but not a sign of interacting galaxy.
This suggests that the host galaxy is a merger remnant and a recoil event could have occurred at the final stage of tidal interaction.

\subsection{M$_{BH} - \sigma$ relation}

The recoil velocity we have measured is only $\sim$10\% of the typical escape velocity in elliptical galaxy ($v_{\rm e}\approx$ 1500 - 2000 km s$^{-1}$).
In such a case, the rSMBH will undergo damped oscillations around the center of host galaxy (Gualandris \& Merritt 2008; Blecha et al. 2011).
The mass of the oscillating SMBHs may be up to five times less massive than their stationary counterparts
and could be a source of intrinsic scatter in the SMBH and stellar bulge mass scaling law (Blecha et al. 2011).
As shown in Fig. 3, our previous study (Kim et al. 2016) supports their claim:
the black hole mass (M$_{BH}$) of the kinematically-identified rSMBH candidates (blue circles) are
on average 5.2$\pm$3.2 times smaller than their SDSS stationary counterpart (green dots). 
The magenta circle in Fig. 3 is the data point of the CXO J101527.2+625911 whose black hole mass M$_{BH}$ = 10$^{8.21\pm 0.02}M_\odot$ was calculated by the virial method (Ho \& Kim 2015)
and velocity dispersion $\sigma_*$ (190$\pm$20 km s$^{\rm -1}$) was estimated from line width of [S II]6716 (Komossa \& Xu 2007).
It is interesting to notice that the CXO J101527.2+625911 does not fall in this scenario and 
more closely follows M$_{BH}-\sigma_*$ correlation found in the ellipticals and classical bulges (solid line: Kormendy \& Ho 2013).
The weak tidal feature, lack of any anomalies in the residual image in Fig. 1d, and small spatial (1.25 kpc) and velocity (175 \kms) offsets
may suggest that the CXO J101527.2+625911 is in the final stage of damped oscillations and the whole system could turn into the normal elliptical.
The Eddington ratio and M$_{BH}$ of the CXO J101527.2+625911 are about 6 times smaller and 6 times larger than those of
rSMBH candidates (Fig. 3b), respectively, but are similar to values found in X-ray selected broad-line AGNs (Suh et al. 2015).
This could indicate that it has almost finished growing its bulge and black hole masses via accretion.

\subsection{Star Formation Rate}
It is suggested that star formation in galaxies is regulated by AGN feedback and outflow (Fabian 2012; Tombesi et al. 2015).
If the central AGN is displaced by a recoil event, it will no longer result in quenching but instead enhance central star formation in the host galaxy (Blecha et al. 2011; Sijacki et al. 2011).
The infrared luminosity of CXO J101527.2+625911 calculated from IRAS ADDSCAN/SCANPI values is log (L$_{FIR} = 1.6\times 10^{12}$ L$_{\odot}$), which
puts this galaxy into an ultraluminous infrared galaxy category (ULIRG: L$_{FIR} > 10^{12}$ L$_{\odot}$).
The large infrared luminosity supports the idea of enhanced star formation in this AGN-displaced host galaxy.
The star formation rates (SFRs) in CXO J101527.2+625911 estimated from 
infrared luminosity (Kennicutt \& Evans 2012), 1.4 GHz radio luminosity (Murphy et al. 2011), 
and H$\alpha$+24 $\mu m$ luminosities are 240 M$_\odot$yr$^{-1}$, 387 M$_\odot$yr$^{-1}$, and 128 M$_\odot$yr$^{-1}$, respectively.
Predicted SFRs from simulations (Blecha et al. 2011; Sijacki et al. 2011) in AGN-displaced host galaxies are less constrained and range from a few M$_\odot$yr$^{-1}$
to a few $\times\ 10^3$ M$_\odot$yr$^{-1}$ and our estimated value fits in the middle of the predictions.
However, unlike in the kinematically identified rSMBH candidates (Kim et al. 2016) 
we do not detect Wolf-Rayet features in the spectra
suggesting no recent star formation activity has happened in this galaxy.
On the other hand, we find a high-excitation coronal line of [Ne V]$\lambda$3426 ($I_p$=97.11 eV)
which is an unambiguous sign of AGN activity.

\section{Summary}
A systematic imaging and spectroscopic search for rSMBHs was undertaken and resulted a detection of a potential rSMBH candidate CXO J101527.2+625911.
The following summarizes our findings:

$\bullet$
A spatially offset (1.26$\pm$0.05 kpc) nucleus and redshifted (175$\pm$25 km s$^{\rm -1}$) H$\beta$ broad-line were detected in the CXO J101527.2+625911.

$\bullet$
A dual SMBHs scenario was investigated.
The column density towards the nucleus of the host galaxy was found to be Compton thin, but no X-ray source was detected.
The null detection of the SMBH in the nucleus in the Chandra observation
suggests either the SMBH does not exist or it exists but has a low accretion rate (LLAGN).
However, the existence of LLAGN was found to be unlikely based on the 
L$_{H\alpha}$, P$_{radio}$, and orbital velocity arguments.
The X-ray and LLAGN arguments favor a rSMBH scenario.

$\bullet$
The host galaxy is a bulge-dominated elliptical and shows a weak morphological disturbance in the outskirts of the galaxy, suggesting a post merger.

$\bullet$
The SFRs in CXO J101527.2+625911 estimated from infrared, radio, and H$\alpha$+24 $\mu m$ luminosities 
range from 128 M$_\odot$yr$^{-1}$ to 387 M$_\odot$yr$^{-1}$.

$\bullet$
The black hole mass (log M$_{BH}$ = 8.21$\pm$0.02 M$_\odot$) and Eddington ratio (0.09$\pm$0.01) of the CXO J101527.2+625911 are similar to those found in normal ellipticals.
The small spatial and velocity offsets, weak morphological disturbance, elliptical galaxy type, and normal black hole mass and Eddington ratio similar to that of elliptical galaxies
suggest that it has nearly completed accretion and black hole mass growth and is about to turn into a normal elliptical.

 
\noindent
\section{Acknowledgements}

\noindent
The authors thank the anonymous refree for comments and suggestions that greatly improved this paper.
We also thank T. Treu and J.-H. Woo for sharing their reduced Keck spectra and sending comments of the manuscript and S.D. Kim and D.S. Kim for proofreading.
This research has made use of the NASA/IPAC Extragalactic Database
(NED) which is operated by the Jet Propulsion Laboratory, California
Institute of Technology, under contract with the National Aeronautics and Space Administration. 
The scientific results reported in this article are also based in part on data obtained from the Chandra Data Archive.
D.K., I.Y., A.E., and S.S acknowledge support from the National Radio Astronomy Observatory (NRAO) and G.C.P. was supported by a FONDECYT Postdoctoral Fellowship (No.\ 3150361). 
The National Radio Astronomy Observatory is a facility of the National Science Foundation operated under cooperative agreement by Associated Universities, Inc.

\begin{deluxetable}{lcccccccc}
\tabletypesize{\small}
\tabletypesize{\scriptsize}
\tablewidth{0pt}
\tablecaption{Properties of the CXO J101527.2+625911}
\tablehead{
\multicolumn{1}{c}{z} &
\multicolumn{1}{c}{Spatial offset} &
\multicolumn{1}{c}{Velocity offset} &
\multicolumn{1}{c}{${L_{0.5-7.0 keV}}$} &
\multicolumn{1}{c}{${L_{1.4 GHz}}$} &
\multicolumn{1}{c}{${L_{FIR} \over L_\odot}$} &
\multicolumn{1}{c}{$\sigma_v$} &
\multicolumn{1}{c}{${M_{BH} \over M_\odot}$} &
\multicolumn{1}{c}{${L_{bol} \over L_{Edd}}$} \\
\multicolumn{1}{c}{} &
\multicolumn{1}{c}{\arcsec\ \ (kpc)} &
\multicolumn{1}{c}{\kms} &
\multicolumn{1}{c}{log} &
\multicolumn{1}{c}{log} &
\multicolumn{1}{c}{log} &
\multicolumn{1}{c}{\kms} &
\multicolumn{1}{c}{log} &
\multicolumn{1}{c}{}\\
\multicolumn{1}{c}{(1)} &
\multicolumn{1}{c}{(2)} &
\multicolumn{1}{c}{(3)} &
\multicolumn{1}{c}{(4)} &
\multicolumn{1}{c}{(5)} &
\multicolumn{1}{c}{(6)} &
\multicolumn{1}{c}{(7)} &
\multicolumn{1}{c}{(8)} &
\multicolumn{1}{c}{(9)} 
}
\startdata
0.3504 & 0\farcs26$\pm$0\farcs01(1.26$\pm$0.05) & $175\pm 25$ & 43.38 & 23.79 & 12.20 & $ 190\pm 20 $ & $8.21\pm 0.02$ & $ 0.09\pm 0.01$ \\
\enddata

(1) Redshift. (2) Projected spatial offsets in arcsec and kpc units. (3) Line of sight velocity offset from H$\beta$ broad-line. (4) Chandra 0.5-7 keV X-ray luminosity in ergs s$^{-1}$ cm$^{-2}$ unit. (5) FIRST 1,4 GHz radio luminosity in Watts Hz$^{-1}$ unit. (6) Far-Infrared luminosity. (7) Velocity dispersion. (8) Black hole mass estimated from virial method. (9) Eddington ratio calculated from virial black hole mass.

\end{deluxetable}

\clearpage

\begin{figure}
  \centering
  \begin{overpic}[scale=0.8]{fig1a.eps}
     \put(1,5){\includegraphics[scale=0.90]{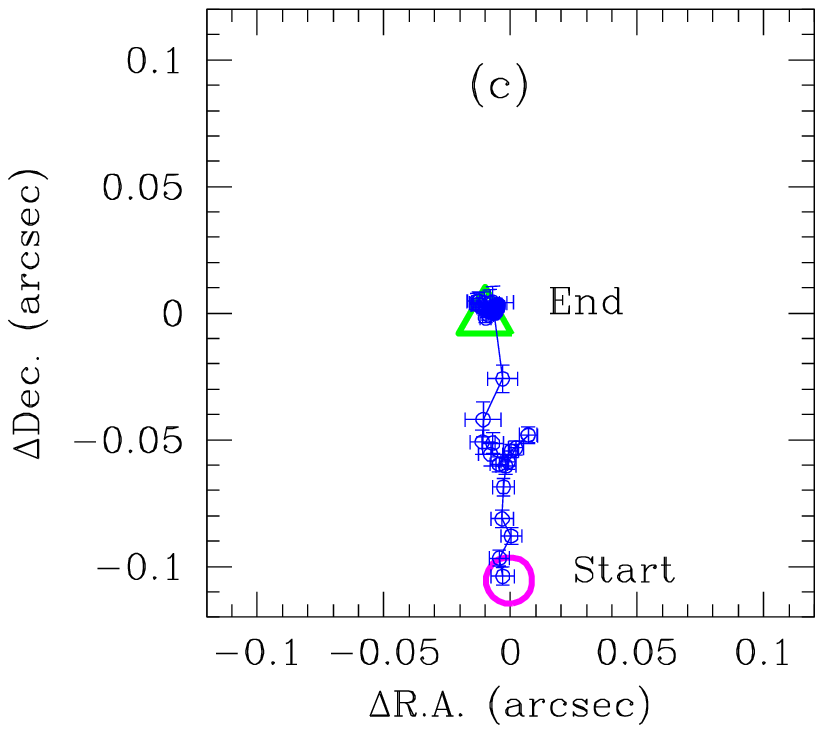}} 
  \end{overpic}
\caption{(a) The HST ACS image of CXO J101527.2+625911,
(b) Zoom-in image of the boxed region in panel (a),
(c) Positional offsets between ellipse fitting centers and southern nucleus, and
(d) Model galaxy used in Galfit (upper-left: psf, upper-right: bulge, lower-left: disk, lower-right: residual (data - model)).
The white dot and circle in panel (b) represent the LQRF quasar position and its positional uncertainty, and
the magenta plus symbol and circle represent location of the Chandra x-ray source and its positional uncertainty.
Contour levels are spaced in log (5) units apart.
The horizontal bar in panel (a) represents 10 kpc physical scale.
North is up and east is to the left.
}

\end{figure}


\begin{figure}[!hb]
\centerline{\includegraphics[scale=0.8]{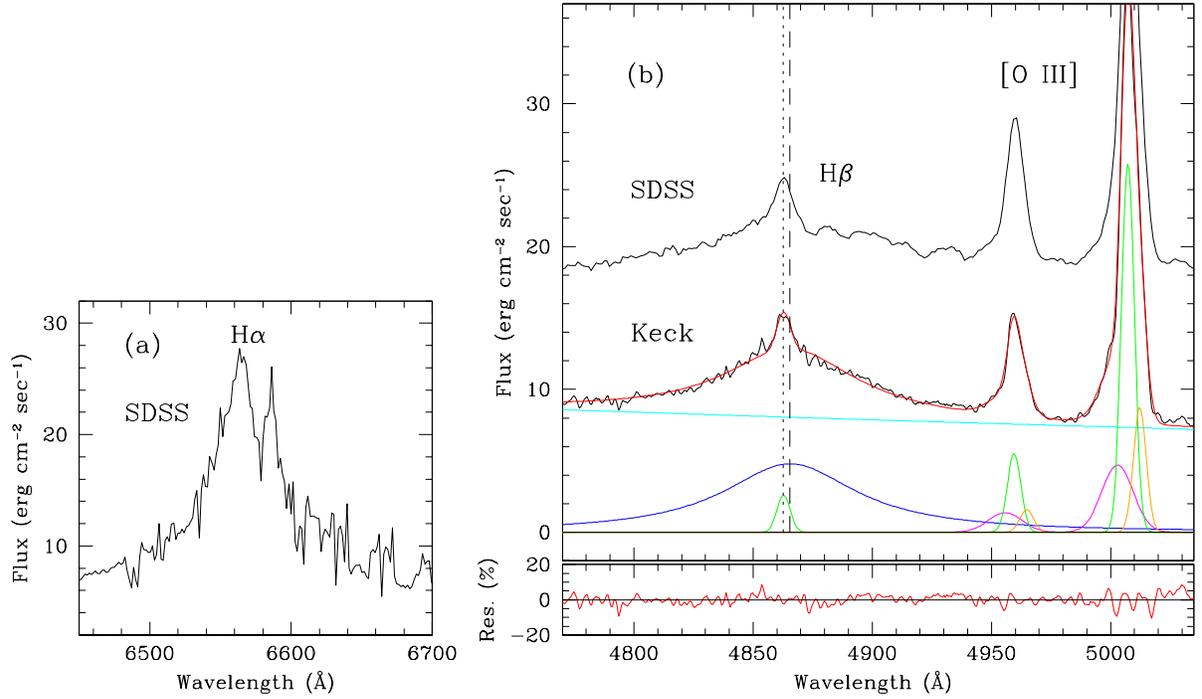}}
\caption{(a) SDSS spectra near H$\alpha$ line, (b) Result of spectral decompositions of the H${\beta}$ line.
Dotted and dashed lines represent line centers of the H$\beta$ systemic velocity and the H$\beta$ broad-line, respectively.
The residual of the fitting (data/model in percentage) is shown on the bottom of the plot.
The SDSS spectrum is shown for comparison.
}

\end{figure}

\clearpage
\begin{figure}[!ht]
\centerline{\includegraphics[scale=0.8]{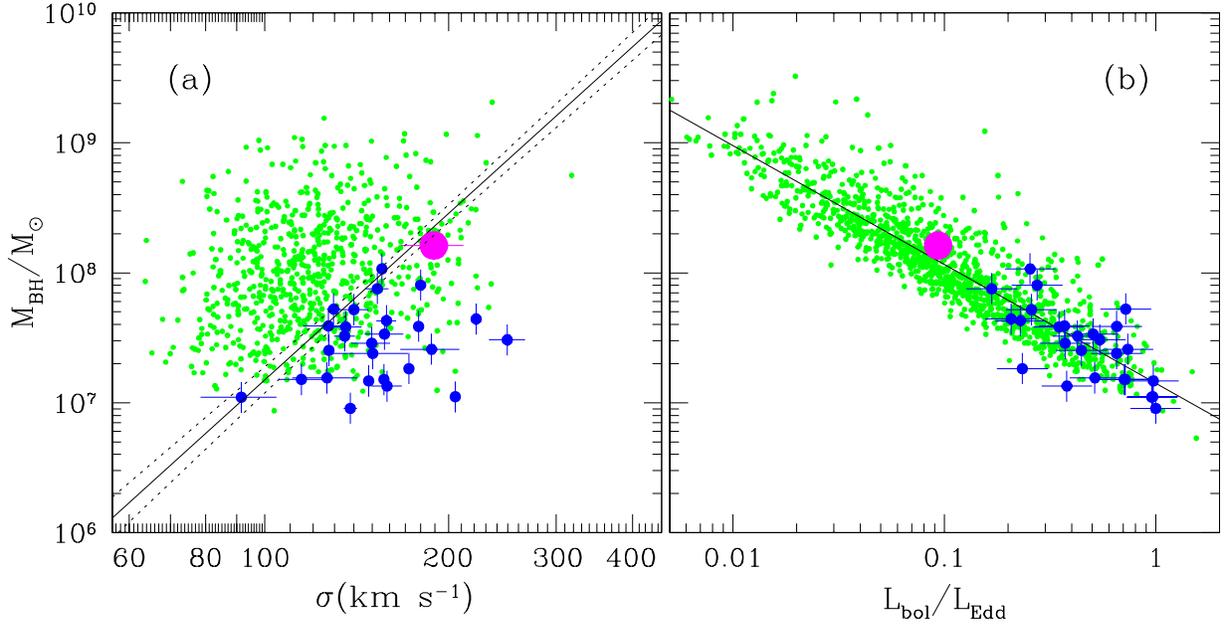}}
\caption{(a) M$_{BH}-\sigma_*$ relation and (b) M$_{BH}$ vs. Eddington ratio plots.
Blue circles and green dots represent kinematically identified rSMBH candidates and
their stationary counterpart of SDSS QSOs with z$<$ 0.25, respectively.
The solid and dotted lines on the left panel represent least square fit for ellipticals and classical bulges and its 1$\sigma$ scatter, respectively.
The solid line of the right panel represents least square fit for the SDSS QSOs.
Magenta circle represents data point of CXO J101527.2+625911.
}

\end{figure}

\clearpage

\end{document}